\documentclass[%
 reprint,
superscriptaddress,
 amsmath,amssymb,
 aps,
]{revtex4-1}

\usepackage{graphicx}
\usepackage{dcolumn}
\usepackage{bm}
\usepackage{times}

\begin{document}


\title{Thermodynamic signatures of chain segmentation in dilute quasi-one dimensional Ising systems} 

\author{Johnathan M. Bulled}
\affiliation{Inorganic Chemistry Laboratory, University of Oxford, South Parks Road, Oxford OX1 3QR, U.K.}%
\author{Mario Falsaperna}
\affiliation{School of Chemistry and Forensic Science, University of Kent, Canterbury CT2 7NH, U.K.}%
\affiliation{Department of Chemistry, University of Bath, Claverton Down, Bath, BA2 7AY, U.K.}%
\author{Paul J. Saines}
\affiliation{School of Chemistry and Forensic Science, University of Kent, Canterbury CT2 7NH, U.K.}%
\author{Andrew L. Goodwin}
\affiliation{Inorganic Chemistry Laboratory, University of Oxford, South Parks Road, Oxford OX1 3QR, U.K.}%

\date{\today}

\begin{abstract}
Heat-capacity measurements are a useful tool for understanding the complex phase behaviour of systems containing one-dimensional motifs. Here we study the signature within such measurements of the incorporation of defects into quasi-one-dimensional (q-1D) systems. Using Monte Carlo  simulations, we show that, on dilution by non-interacting sites, q-1D Ising models display a low-temperature signature not present in conventional three-dimensional models. Frustrated embeddings of 1D chains show similar features to unfrustrated embeddings, with the additional emergence of a finite ground-state entropy in the former, which arises from chain segmentation. We introduce a mean-field formulation which captures the low-temperature behaviour of the model. This theoretical framework is applied to the interpretation of experimental heat-capacity measurements of the Tb$_{1-\rho}$Y$_{\rho}$(HCOO)$_3$ family of magnetocalorics. We find good correspondence between simulation and experiment, establishing this system as a canonical example of dilute q-1D magnets. 
\end{abstract}

\maketitle

\begin{figure}
	\begin{center}
		\includegraphics[width=\linewidth]{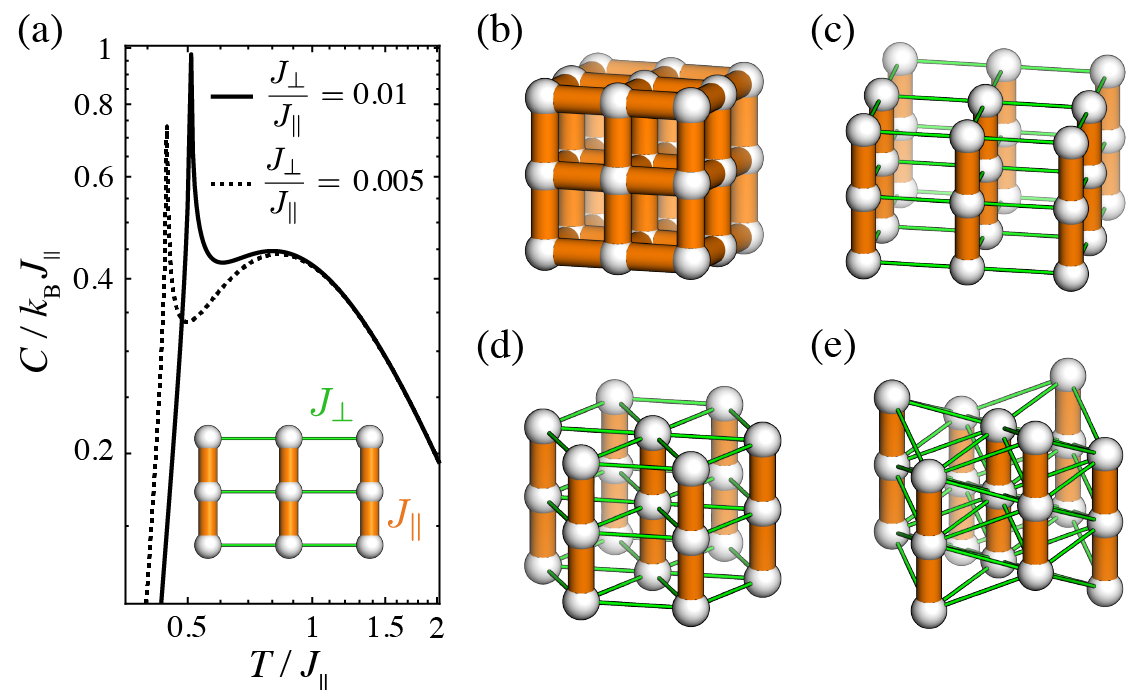}
	\end{center}
	
	\caption{\label{fig1}(a) Onsager's exact solution \cite{Onsger_1944} for the heat capacity of an anisotropic rectangular lattice, shown for $J_\perp/J_\parallel =0.01$ and 0.005, is very sensitive to the ratio of coupling constants. A fragment of the lattice is shown as an inset, with intra-chain ($J_\parallel$) and inter-chain ($J_\perp$) interactions represented by (thick) orange and (thin) green rods respectively. (b--e) Lattices considered in this paper: (b) the 3D cubic lattice; (c) the q-1D square lattice; (d) the q-1D triangular lattice; (e) the staggered triangular terbium sublattice of the $R3m$ Tb(HCOO)$_3$ crystal structure \cite{Falsaperna_2022}.
 }
\end{figure}

The existence of partial order has been well characterised in a range of quasi-one dimensional (q-1D) systems---materials containing one dimensional (1D) motifs embedded in a weakly interacting three dimensional (3D) lattice; examples include urea clathrates \cite{Welberry__1996,Simonov_2022}, charge-density wave phases \cite{Lynn_1975}, organic conductors \cite{Bloch_1974}, inorganic and molecular magnets \cite{Landee_1990,Dingle_1969,Coulon_2006,Harcombe_2016}, and ferroelectrics \cite{Kumari_2022,Furukawa_2021,Jangen_2005}. These systems are an important playground for theoretical study \cite{Glauber_1963, Ising_1925, Faddeev_1981}, because they exhibit a wealth of unusual physics \cite{Nagler_1991, Tennant_1993}. A key signature of q-1D physics lies in the temperature dependence of the heat capacity $C(T)$: the function contains a broad feature associated with the development of order within 1D motifs, and a lower-temperature peak signifying the onset of long-range 3D order [Fig.~\ref{fig1}(a)]. Small changes in the strength of inter-chain coupling give rise to large changes in $C(T)$, as seen in Onsager's analytical solution to the heat capacity of the anisotropic rectangular lattice [Fig.~\ref{fig1}(a)]. This sensitivity is why heat capacity measurements are such an effective tool for model parameterisation of q-1D systems \cite{Landee_1990}.

Our interest here is in understanding the role of defects in modifying q-1D physics. Because the 1D lattice is unique in having no percolation threshold---a single vacancy divides the lattice in two---q-1D systems have been found to be surprisingly sensitive to small defect concentrations \cite{Coulon_2006}. 
For example, defect pinning and incommensurations play an accidental role in the physics of 1D charge density wave (CDW) materials \cite{Teranashi_1979, Hamano_1985}. Likewise, the deliberate inclusion of nonmagnetic defects is emerging as an effective strategy both to tune magnetocaloric effects \cite{Orendac_2020, Falsaperna_2022} and to develop unconventional ferromagnets \cite{Wang_2021}. 
Given the importance of thermodynamic measurements in understanding q-1D materials, an obvious outstanding question is how the incorporation of defects affects $C(T)$.

To address this question, we focus on the well-studied problem of nonmagnetic doping (dilution) of Ising models \cite{Griffiths_1969,Wortis_1974, Harris_1975, McCoy_1968}. To quantify the roles of dimensionality and geometric frustration, and to allow direct comparison against experiment in due course, four simple models are considered [Fig.~\ref{fig1}(b--e)]. We are chiefly  interested in the family of q-1D models involving chains of ferromagnetically coupled Ising \mbox{(pseudo-)spins}, which interact via a much weaker antiferromagnetic inter-chain coupling. Such models have the Hamiltonian
\begin{equation}\
\mathcal H = -J_\parallel\sum_{\left<i,j\right>_\parallel} o_i o_j {S}_i  {S}_j + J_\perp \sum_{\left<i,j\right>_\perp} o_i o_j {S}_i  {S}_j,
\end{equation}
where $J_\parallel\gg J_\perp>0$ are the intra- and inter-chain coupling strengths and the corresponding sums are taken over nearest-neighbour intra- and inter-chain pairs, respectively. The pseudospin variables $ S_i = \pm1$ represent a general Ising state and the corresponding occupancies $o_i$ denote whether the site either has a degree of freedom ($o_i=1$) or is a non-interacting defect ($o_i=0$). The $o_i$ are distributed randomly amongst all sites $i$ subject to an overall defect density $\rho=1-\langle o\rangle$ and are considered immobile---the so-called quenched limit \cite{Coulon_2006}. 
For each of the models studied, it is already well established that dilution leads to a ``Griffiths'' phase, where statistically-rare regions of complete order make such models impossible to solve analytically \cite{Griffiths_1969,Wortis_1974, Harris_1975}.

In this Letter, we apply both computational and theoretical  approaches to study the effects of dilution on q-1D Ising systems: chain-move Monte Carlo (MC) simulations identify a dimensional crossover spanning three temperature regimes, and a mean-field theory (MFT) is used to study the low- and medium-temperature regimes. We identify three thermodynamic signatures of dilute q-1D physics which distinguish it from the behaviour of higher-dimensional systems: (i) the entropy associated with the high-temperature transition decreases linearly with defect concentration; (ii) a maximum emerges at low temperature in the specific heat, which grows in intensity and decreases in temperature with increasing defect concentration; and (iii) below this maximum, the heat capacity decays slowly, with a range where $C(T) \propto T$ emerging at low levels of dilution. We demonstrate that magnetic frustration leads to a broadening of these features and a zero-point-entropy that varies linearly with $\rho$. The theoretical framework is then applied to interpret recent experimental heat capacity measurements of the q-1D magnetocalorics Tb$_{1-\rho}$Y$_{\rho}$(HCOO)$_3$.

\begin{figure}
	\includegraphics[width=\linewidth]{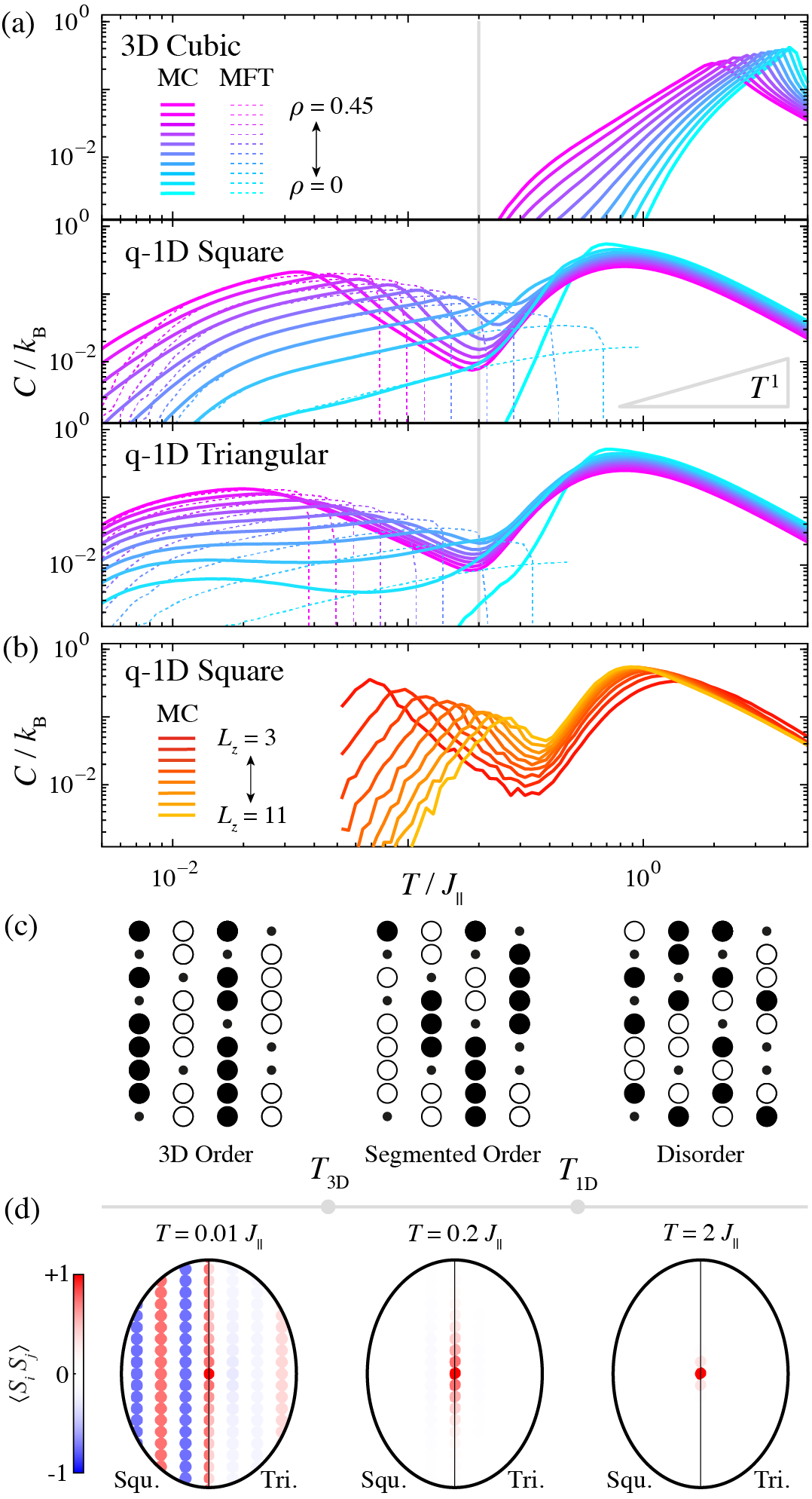}
	\caption{\label{fig2} (a) Comparison of heat capacity per spin in the 3D cubic, q-1D square and q-1D triangular models at ten evenly-spaced vacancy concentrations between $\rho=0$ and 0.45. In each of the q-1D models, $J_\perp/J_\parallel = 10^{-2}$. Solid lines display MC results, and the mean-field theory results are shown as dashed lines.
The two dominant features in the q-1D models are denoted by $T_{\rm 3D}$ and $T_{\rm 1D}$ at low- and high-temperatures, respectively. 
Linear scaling corresponds to the slope of the gray triangle. (b) The effect of finite size on $8\times8\times L_z$ supercell MC simulations of the undoped q-1D square lattice. (c) Schematic showing the three temperature regimes relevant to the physics of dilute q-1D systems:  at $T>T_{\rm 1D}$ the model is disordered; for $T_{\rm 3D}<T<T_{\rm 1D}$ individual chain segments are ordered with weak inter-chain order, and at $T<T_{\rm 3D}$, order develops between segments. Large filled and hollow represent $\pm1$ Ising states while smaller dots represent non-interacting sites. (d) The corresponding spin--spin correlations functions of the $\rho = 0.25$ q-1D square model, calculated perpendicular to the chain axis at three temperatures corresponding to the schematics in (c).  }
	
\end{figure}

\textit{Chain-move MC simulations---} MC simulation of q-1D systems is made difficult by the range of energy scales involved. Conventional algorithms are easily trapped in local minima, since traversing the energy landscape involves overcoming strong intra-chain interactions. One workaround, used \emph{e.g.}\ to model the frustrated q-1D magnet Ca$_{3}$Co$_{2}$O$_{6}$, is the Wang--Landau algorithm \cite{Wang_2001,Yang_2014,Hu_2013}; a second is the application of non-local moves, such as the `loop moves' of spin-ice simulations \cite{Melko_2001,Otsuka_2015,Wang_2012}. We employ a similar non-local approach here, allowing occasional collective `chain moves' of any spin-chain segment (a set of spins in the same chain, uninterrupted by a vacancy). The ratio of accepted chain flips to spin flips varies as a function of temperature \cite{SI}: at low temperatures chain flips dominate in the q-1D systems since flipping a chain segment does not require overcoming the stronger intra-chain interaction. The heat capacity is extracted as the derivative of the internal energy, and the entropy is integrated relative to the high-temperature limit ($T = 100J_\parallel$). 

\textit{Effect of dilution on heat capacity---} Our results [Fig.~\ref{fig2}(a)] show that dilution impacts the specific-heat of the 3D simple-cubic model very differently to the two q-1D models investigated. 
In the cubic model, we see a similar result to that noted in previous studies: namely, that doping acts both to suppress the peak in $C(T)$ associated with 3D ordering to lower temperature and at the same time to broaden this feature \cite{Griffiths_1969,McCoy_1968}. This behaviour is straightforwardly explained: dilution decreases the average number of interacting neighbour-pairs and also makes the distribution of interacting pairs more varied.

By contrast, the effect of dilution on the $C(T)$ functions of q-1D models is more complex. The broad high-temperature feature associated with 1D order is surprisingly resilient to doping, varying only in intensity but always resembling the behaviour expected for isolated 1D chains. More obviously, a low temperature feature appears, which grows in size and is suppressed to lower temperature with dilution. This feature is qualitatively similar in both square and triangular q-1D models, but appears at lower temperature in the frustrated case.  We denote the temperature maxima associated with these two features $T_{\rm 1D}$ and $T_{\rm 3D}$, respectively.

Examination of the individual spin-configurations of the simulation \cite{SI} and spin--spin correlation functions reveal the origin of this signature [Fig.~\ref{fig2}(c,d)].  The two heat capacity features demarcate three temperature regimes: (i) at high temperature only short range correlations exist along the chains, and the system is disordered; (ii) at temperatures between the two $C(T)$ maxima, spins order within individual chain segments but inter-chain order is short-range (iii) at low temperatures, full 3D order of the spins is observed.

Since in the intermediate temperature regime, we expect a q-1D system to act as weakly interacting, ordered chains, the distribution of segment lengths will be central to the underlying physics.  If the vacancies are randomly arranged, this distribution can be  expressed as the probability of a randomly selected site falling into a chain segment of length $\ell$: 
\begin{equation}
P(\ell) = \ell \rho^2 (1-\rho)^\ell.
\end{equation}
The average chain length $\left<\ell \right> = 1 / \rho$: as vacancy concentration is increased, the average chain become shorter. Since the effective interactions between two segmented chains increases with the segment length, it makes sense that the strength of interactions, and therefore the 3D ordering temperature, decreases with increasing vacancy concentration. 

Similar behaviour emerges as a finite-size effect in undoped q-1D models  \cite{Lee_2002}. In our simulations of the $\rho=0$ q-1D square model, varying the size of the system in the chain direction ($L_z$) leads to a substantial change in the heat-capacity [Fig.~\ref{fig2}(b)]. With decreasing $L_z$, a new feature emerges on the low-temperature side of the main $C(T)$ peak, growing in intensity and decreasing in temperature in a way that mirrors the behaviour observed on doping. The two effects are similar because increasing dopant concentration and reducing model size both decrease the average segment length. The main difference between these responses is that the low-temperature feature in Fig.~\ref{fig2}(a) is much broader than that in (b) because there is a distribution of segment lengths in the randomly-doped samples, whereas varying the box size leaves all chain lengths the same.

\textit{Mean-field theory---} This mapping between dilution and finite size effects suggested an approximate formulation for the doped q-1D case. Let us assume that below the 1D ordering transition, all the chain segments in each of $n$ sublattices are internally ordered and interact via a mean-field of their neighbours. To compare to square and triangular lattice models we use bipartite ($ \alpha\in\{1,2\} $) and tripartite ($ \alpha\in\{1,2,3\} $) models, respectively. In this formulation, the Hamiltonian $\mathcal H_{\rm MF}^{\alpha}$ of sublattice $\alpha$ reduces to a sum over segment lengths, where each segment $j\in (\ell, \alpha)$ (of length $\ell$ on sublattice $\alpha$) has a collective normalised spin $S_j = \pm 1$:
\begin{equation}\label{MF_ham}
\mathcal H_{\rm MF}^{\alpha} =  \left[\frac{J_\perp}{2} \sum_{\beta}  Z_{\alpha \beta} M_\beta \right] \left[ \sum_{\ell =1}^\infty   \ell \sum_{j\in (\ell, \alpha)} S_j  \right].
\end{equation}
The first bracket encodes the mean-field of the neighbouring sublattice(s): $Z_{\alpha \beta}$ is the per-site coordination number ({via} inter-chain bonds) between the $\alpha$ and $\beta$ sublattices in the undoped system; $M_\beta$ is the magnetisation of sublattice $\beta$. The second bracket encodes the net spin of sublattice $\alpha$.
The self-consistency equations
\begin{equation}\label{MF_const}
M_\alpha = \sum_{\ell =1}^\infty  P(\ell) \tanh \left[  \frac{\ell J_\perp}{T} \left(\sum_{\beta}  Z_{\alpha \beta} M_\beta  \right) \right]
\end{equation}
follow from \eqref{MF_ham} and can be solved numerically for the bipartite and tripartite cases \cite{SI}. We subsequently obtain the heat-capacity as the derivative of the internal energy. The sum in \eqref{MF_const} converges for all values of $\rho$, and we find a maximum value of $\ell = 1000$ ensures numerical accuracy.

The theory reproduces well the low-temperature behaviour of the square q-1D system, particularly at low vacancy concentrations [Fig.~\ref{fig2}(a)]. It is prone to many of the shortfalls of mean-field theories: the transition temperature is overestimated and the behaviour near the ordering transition is less well captured. The behaviour of the q-1D triangular case is less well captured because geometric frustration leads to a distribution of local spin environments at the lowest temperatures, making the tripartite approximation of this lattice less effective than the bipartite approximation in the square case. Nevertheless, the important point is that the mean-field model captures the dependence of the low-$T$ feature on $\rho$: the transition temperature decreases as $\rho$ increases and the feature at $T_{\rm 3D}$ becomes more pronounced. 

\textit{Linear regime---} It is apparent from the MC results that for low dilution, the square q-1D lattice has a regime where $C(T)$ varies linearly with $T$. This regime spans the largest temperature range for low vacancy concentrations and occurs at a lower temperatures with increased values of $\rho$. In this limit, we can use the mean-field framework developed above to show that there is a regime below $T_{\rm 3D}$ (specifically, $ 1\ll T /2J_\perp Z_\perp\ll {1}/{\rho} $) for which the heat capacity follows the relation
\begin{equation}\label{t2}
    C(T)  = \frac{\gamma \pi^2  \rho^2 k_{\rm B}}{12 J_\perp Z_\perp } T.
\end{equation}
 Here, $\gamma$ is a geometric factor equal to 1 or $\frac{4}{3}$ for bipartite or tripartite lattices, respectively \cite{SI}. Although this approximation is valid only within a small window of the function, Eq.~\eqref{t2} illustrates a more general point: namely, that the decay of the heat capacity below the anomaly maximum is very slow. This is because of the broad distribution of segment lengths---and therefore inter-chain interaction strengths---that are involved in developing full 3D order.

\begin{figure}
	\includegraphics[width=\linewidth]{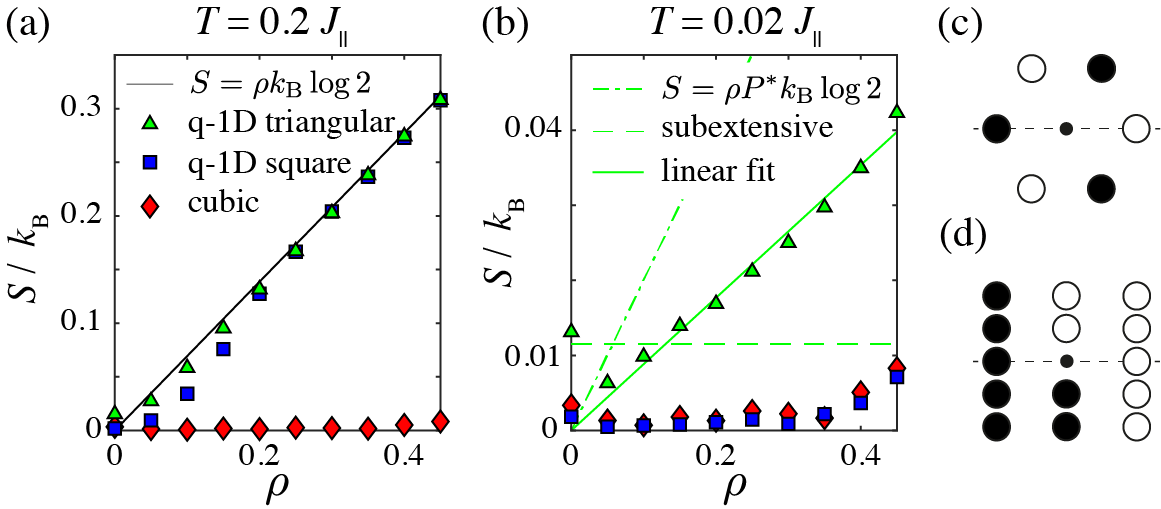}
	\caption{\label{fig3}(a) The intermediate-temperature ($T_{\rm 3D}<T = 0.2 J_\parallel<T_{\rm 1D}$) MC entropy per spin of the three models as a function of dilution. (b) The low-temperature ($T=0.01J_{\parallel}$) entropy, with a linear fit to the q-1D triangular results shown as a solid line and the finite-size limit shown as a dashed line. The upper-bound trend discussed in the text is shown as a dot-dashed line. (c,d) A local state that gives rise to a $\rho$-dependant ground-state entropy, viewed (c) along and (d) perpendicular to the chain axis.}
\end{figure}

\textit{Effect of dilution on entropy---} The entropy is calculated by integrating over the heat-capacity from the high-temperature limit ($T_{\rm max}=100\,J_\parallel$) of our MC simulations (at which point the system is assumed to be maximally disordered):
\begin{equation}
	S(T) = k_{\rm B}\log 2 - \int^{T_{\rm max}}_T \frac{C(T^\prime)}{T^\prime}\,{\rm d}T^\prime.
\end{equation}
In the high-temperature limit of the mean-field theory, chain segments are internally ordered but there exist no correlations between segments. The associated per-spin entropy ($S =  \rho k_{\rm B} \log 2$) is shown in Fig.~\ref{fig3}(a) alongside the values calculated from the MC simulations an intermediate temperature, $T=0.2J_\parallel$, chosen between $T_{\rm 3D}$ and $T_{\rm 1D}$. The relation matches closely for all but the lowest values of $\rho$, where correlations between segments lead to a suppression of entropy. By contrast, the 3D cubic model is completely ordered for all values of $\rho$. 

When evaluated to the low-temperature limit, the entropies of the undoped q-1D models join that of the cubic model, with only minor deviations from complete order [Fig.~\ref{fig3}(b)]. The low-temperature entropy of the triangular model shows a complex evolution as a function of doping. $S$ varies approximately linearly with $\rho$ for all nonzero values of $\rho$, but jumps at $\rho=0$ because of a subextensive entropy contribution that is a finite-size effect. The value of this contribution is given by theory as $S = 0.3231 k_{\rm B} / L_z$, which matches well the simulated value \cite{Wannier_1950}. A small addition of vacancies breaks this degeneracy by creating an imbalance in the geometrically-frustrated interactions between chain segments of varying lengths, which is the reason for the initial decrease in entropy on doping. By contrast, the linear entropy increase with further doping is robust to changes in the simulation size. The empirical relation $S \simeq  0.128\,\rho k_{\rm B} \log 2$ implies a $\sim$12.8\% chance of a vacancy doubling the ground-state degeneracy.  This result contrasts that of higher-dimensional systems such as the spin ices, where doping leads to a non-monotonic variation of residual entropy with dilution \cite{Ke_2007,Lin_2014}. 

Why does residual entropy scale linearly with defect density? To explain this trend, we consider environments such as that shown in Fig.~\ref{fig3}(c,d), where the spins in the central column could take either state with minimum energy. Taken in isolation, the addition of a vacancy of this kind would double the ground-state degeneracy. Using a MC method, we calculate the probability of selecting such an environment in the two-dimensional triangular Ising antiferromagnet as $P^*$ = 0.289(3) \cite{SI}; the corresponding entropy [shown as a dot-dashed line in Fig.~\ref{fig3}(b)] is a severe overestimate because it neglects the entropy-suppressing effect already noted: vacancies not only create environments like (d) but also make the surrounding chain-segments more unequal. Environments like this are unique to geometrically-frustrated lattices, which explains why this effect is not seen in the square or cubic models. 
 
 \begin{figure}
 	\includegraphics[width=\linewidth]{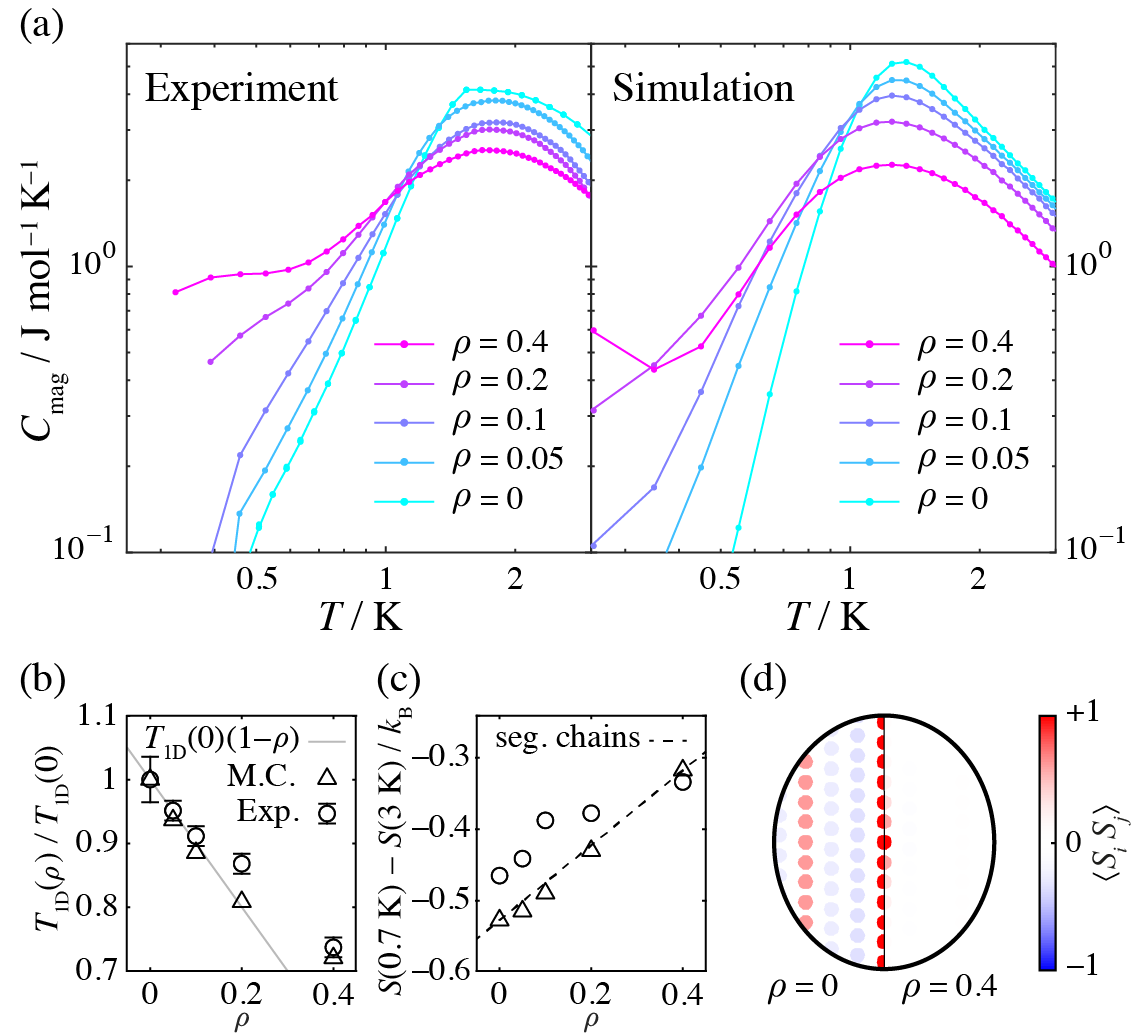}
 	\caption{\label{fig4}(a) Experimental and simulated magnetic heat capacity of the Tb$_{1-\rho}$Y$_\rho$(HCOO)$_3$ series. (b) The variation in $T_{\rm 1D}$ with degree of dilution for simulation (triangles) and experiment (circles), compared with the mean-field result (line). (c) Entropy change over the 1D ordering transition as a function of dilution. Markers have the same meanings as in (a), but the solid gray line is now the simple relation $\Delta S(\rho) = \Delta S(\rho=0)(1-\rho)$, expected for complete in-chain order. (d) Spin--spin correlation functions generated from the $T = 0.4$\,K MC simulations with $\rho = 0,0.4$.}
 \end{figure}
 
\textit{Application to experiment---} We conclude by relating our theoretical and computational results to the experimental behaviour of the q-1D magnets Tb$_{1-\rho}$Y$_\rho$(HCOO)$_3$. In this family, chains of magnetic Tb$^{3+}$ and non-magnetic Y$^{3+}$ ions are arranged on a triangular lattice. Intra-chain interactions are ferromagnetic; inter-chain interactions are weaker and antiferromagnetic. The experimental heat-capacity has recently been measured as a function of doping, and is shown in Fig.~\ref{fig4}(a) \cite{Falsaperna_2022}. We use an Ising variant of the Hamiltonian of Ref.~\citenum{Harcombe_2016} to carry out MC simulations for the experimental values of $\rho$. The results of both experiment and simulation are shown in Fig.~\ref{fig4}(a): the two specific heat functions display a clear separation between a high temperature ordering transition and an upturn at low temperature that increases in intensity as $\rho$ increases. The very low temperatures involved mean that we were unable to access the full 3D ordering transition experimentally, but we interpret our results nonetheless as consistent with theory. Moreover, there is good quantitative agreement between the defect-dependence of the position of the maximum in $C(T)$ [Fig.~\ref{fig4}(b)], and also of the entropy change associated with 1D order [Fig.~\ref{fig4}(c)]. In both simulation and experiment, the entropy-change follows the $\Delta S(\rho) = \Delta S(\rho=0)(1-\rho)$ law, consistent with the linear trend in Fig.~\ref{fig3}(a), so nonmagnetic doping has led to a residual entropy at $0.7$\,K. On examining the spin correlations of our simulation [Fig.~\ref{fig4}(d)], we see that inter-chain correlations are disordered in the $\rho=0.4$ segmented phase to the lowest experimentally-accessible temperature.

{\it Concluding remarks---} So as to study the thermodynamic signature of doping in q-1D systems systematically, we have intentionally limited the scope of our study to the Ising model. Nevertheless, our analysis might in future be extended straightforwardly to XY or Heisenberg degrees of freedom, and/or to include long-range interactions (\emph{e.g.} dipolar)---so long as these additional interactions are much weaker than the intra-chain coupling.  The prediction that $C \propto T$ at low temperature for low concentrations of impurities may be of interest to in quantum magnetism, where linear heat capacities are used as a hallmark of fermionic excitations \cite{Liang_2015} but the presence of defects is not considered.  Such studies may also be relevant to the behaviour of a variety of q-1D CDW systems, for example, where chemical or microstructural defects are thought to play an important role \cite{Hamano_1985}.
A key advantage of exploiting heat capacity measurements as a probe of unconventional order is that there is no fundamental constraint on the type of degree of freedom involved. This is why our analysis might be as relevant to type I CDWs, q-1D ferroelectrics, and other q-1D systems as it is to the behaviour of frustrated magnets. 

\begin{acknowledgments}
The authors thank J. Paddison (Oak Ridge), E. Lhotel (Institut N\'eel), and J. Neilson (Colorado State) for useful discussions, and gratefully acknowledge the E.R.C. (Grant 788144), E.P.S.R.C. (Grant EP/T027886/1) and Leverhulme Trust (Grant RPG-2018-268) for financial support.

\end{acknowledgments}

\end{document}